\documentclass[sn-mathphys,Numbered,twocolumn]{sn-jnl}
\usepackage{graphicx}%
\usepackage{multirow}%
\usepackage{amsmath,amssymb,amsfonts}%
\usepackage{amsthm}%
\usepackage{mathrsfs}%
\usepackage[title]{appendix}%
\usepackage{xcolor}%
\usepackage{textcomp}%
\usepackage{manyfoot}%
\usepackage{booktabs}%
\usepackage{algorithm}%
\usepackage{algorithmicx}%
\usepackage{algpseudocode}%
\usepackage{listings}%
\usepackage{caption}
\usepackage{multicol}
\usepackage{pdfcomment} 
\usepackage{datetime2}
\usepackage{comment}
\usepackage{textcomp}

\begin{document}

\newgeometry{
margin=1in,
}

\title[Article Title]{\textbf{Ferromagnetic interlayer exchange coupling in a few layers of CrSBr on a gold thin film}}

\author[1]{\fnm{Rixt} \sur{Bosma$^{\times}$}} 

\author[1]{\fnm{Darius A.} \sur{Pacurar$^{\times}$}} 

\author[1]{\fnm{Daniel} \sur{Sade}} 

\author[1]{\fnm{Jingbo} \sur{Wang}} 

\author[2,3]{\fnm{Nicholas} \sur{Dale}} 

\author[2]{\fnm{Cameron W.} \sur{Johnson}}

\author[4]{\fnm{Sergii} \sur{Grytsiuk}}

\author[4]{\fnm{Alexander} \sur{Rudenko}}

\author[2]{\fnm{Alexander} \sur{Stibor}}

\author[4,5]{\fnm{Malte} \sur{R\"osner}}

\author[1]{\fnm{Marcos H. D.} \sur{Guimarães}}

\author*[1]{\fnm{Roberto} \sur{Lo Conte}} \email{r.lo.conte@rug.nl}

\affil[1]{\orgdiv{Zernike Institute for Advanced Materials, University of Groningen}, \city{Groningen}, \country{Netherlands}}

\affil[2]{\orgdiv{Molecular Foundry, Lawrence Berkeley National Laboratory}, \city{Berkeley, CA}, \country{USA}}

\affil[3]{\orgdiv{Materials Science Division, Lawrence Berkeley National Laboratory}, \city{Berkeley, CA}, \country{USA}}

\affil[4]{\orgdiv{Institute for Molecules and Materials, Radboud University}, \city{Nijmegen}, \country{Netherlands}}

\affil[5]{\orgdiv{Faculty of Physics, Bielefeld University}, \city{Bielefeld}, \country{Germany}}

\abstract{
The two-dimensional character of van der Waals magnets allows for efficient control of their properties via proximity effects and electrical stimuli, making them promising candidates for application in spin-electronics. We use spin-polarized low energy electron microscopy to directly image the magnetic texture of thin CrSBr on top of a Au film, uncovering a ferromagnetic ground state for CrSBr thicknesses smaller than 11 nm. We argue that the stabilization of the ferromagnetic ordering - as compared to the conventional antiferromagnetic one - is obtained via electron transfer from the Au film to the CrSBr flakes, in agreement with ab-initio density functional theory calculations. Reflected-electron spectroscopy shows clear differences in the unoccupied density of states between a few layers of CrSBr on Au and bulk CrSBr, pointing towards electronic band structure modification in thin CrSBr. This work sheds light on the possibility to tune magnetic properties of two-dimensional magnets via substrate engineering.
}

\maketitle

\vspace*{0.3\baselineskip}

\newpage

\section{Introduction}
Two-dimensional (2D) van der Waals (vdW) magnetic materials are a promising platform for the exploration of magnetism in low dimensions~\cite{BurchNATURE2018,GibertiniNAT-NANOTECH2019}.
Because of their reduced dimensionality, 2D vdW magnets allow for efficient tuning of their magnetic properties through external stimuli. As a result, these materials hold high promise for applications, such as magnetic data storage and logic devices.\newline
Although vdW magnetic materials and the tuning of their properties through electrostatic gating have been studied extensively from an experimental point of view\cite{DengNATURE2018,SharpeSCIENCE2019,TabatabaVakili2024,Hendriks2024,YaoNAT-NANOTECH2025}, the impact of the substrate on their magnetic properties has been much less explored experimentally, despite several theoretical predictions~\cite{SorianoNPJ-COMP-MATER2021,BianchiPRB2023,Rudenko2023,XieACS-NANO2023}.
For applications in actual devices this is of particular importance since the substrate and electrical contacts could locally influence the properties of the system, resulting in magnetic ground states which differ from what is expected for pristine and free-standing 2D magnets. Additionally, these interactions could be used to tune material properties~\cite{BianchiPRB2023,WatsonNPJ-2DmaterAppl2024}, leading to local control of magnetism through substrate engineering. \newline
The prototypical vdW magnet CrSBr represents a promising candidate for spintronic applications. CrSBr is particularly interesting due to its decent air-stability and relatively high magnetic ordering temperature of 132 K \cite{Telford2020}. It was shown that the magnetic ordering of CrSBr can be tuned from the A-type antiferromagnetic ordering to the ferromagnetic ordering via ions intercalation~\cite{ZhaiPNAS2026} or electric gating~\cite{TabatabaVakili2024}, due to a modification of the interlayer magnetic exchange. Previous studies based on band structure mapping by angular-resolved photo-electron spectroscopy (ARPES) offered an initial understanding of how the electronic and magnetic properties of CrSBr are modified by the proximity to metallic substrates~\cite{BianchiPRB2023,WatsonNPJ-2DmaterAppl2024,Ghimirey-npj2DMAT_APPL2026}. Furthermore, the imaging of the magnetic ground state of thin CrSBr flakes on substrates have also been performed, mainly via nitrogen-vacancy (NV) center microscopy~\cite{Tschudin2024,GhiasiNPJ-2DMAT-APP2023}, magnetic force microscopy~\cite{RizzoADV-MAT2022}, and SQUID on tip microscopy~\cite{ZurADV-MAT2023}. However, as of today, direct imaging of the modification of the magnetic ground state of few-layer CrSBr flakes via interface effects is still missing. \newline
Here we report on the magnetic imaging of few-layer CrSBr flakes via spin-polarized low-energy electron microscopy (SPLEEM). We discover a ferromagnetic ground state in few monolayers-thick CrSBr flakes in direct proximity to a metallic gold (Au) film on a SiO$_{x}$/Si substrate. Furthermore, a modification of the electronic band structure of the thin CrSBr flake in proximity to the Au film is detected via spatially-resolved reflected electron spectroscopy (RES). The emergence of a ferromagnetic interlayer exchange coupling in CrSBr is analyzed through first-principles density functional theory calculations, which suggest charge transfer at the CrSBr-Au interface as a likely explanation for the observed ferromagnetic ground state.

\section{Results}

\begin{figure}
    \centering
    \includegraphics[width=6.4in]{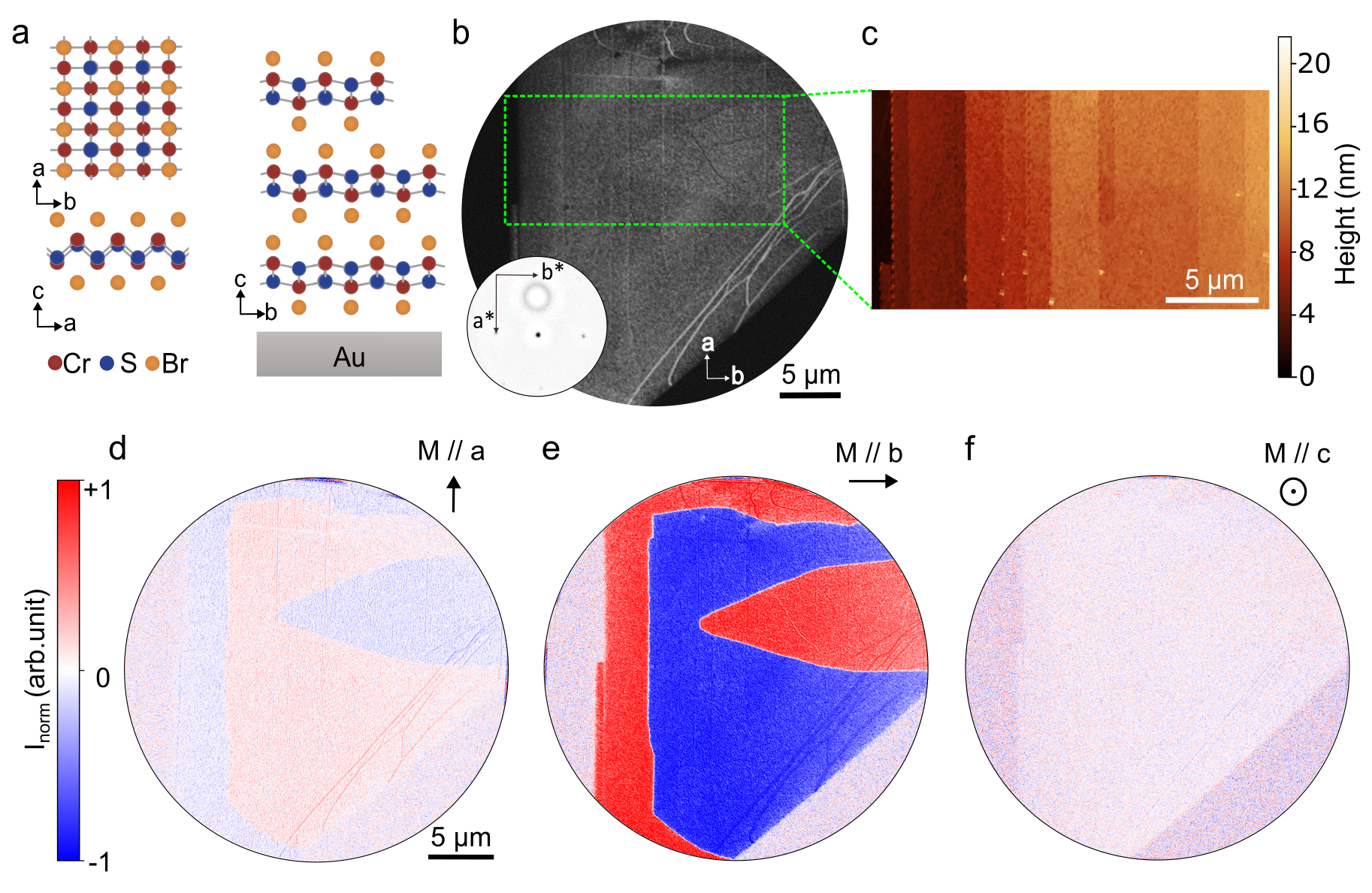}\caption{Schematic of the prepared CrSBr-flake/Au(20 nm)-thin film sample and its structural and magnetic characterization. \textbf{a} The CrSBr flake has an orthorhombic crystal structure. The material is placed on a polycrystalline Au film. \textbf{b} Low-energy electron microscopy (LEEM) image of the investigated flake. The central area of this flake is imaged with atomic force microscopy (AFM) and displayed in \textbf{c}. The inset shows the measured low-energy electron diffraction pattern at a bias voltage of 31.6 V. \textbf{d-f} Spin-polarized LEEM (SPLEEM) images of the CrSBr flake with spin polarization along the three main crystallographic directions, acquired with a bias voltage of 7.5 V. \textbf{d,e} SPLEEM images of the same CrSBr flake in \textbf{b} taken with spin polarization approximately along the \textit{a}- and \textit{b}-axis, respectively. The SPLEEM contrast is maximal in \textbf{e}, consistent with a magnetic easy axis parallel to the \textit{b}-axis. \textbf{f} SPLEEM image for spin polarization along the \textit{c}-axis, showing no magnetic contrast.}
    \label{fig:1}
\end{figure}

CrSBr crystallizes into an orthorhombic structure which comprises double layers of chromium sulfide, capped on both sides by an anionic bromide layer. Monolayers of CrSBr are hold together by weak vdW forces. Along each of the three main crystallographic directions, the exchange interaction between the neighboring magnetic Cr atoms is different, defining the magnetic ordering in the CrSBr crystal. Pristine CrSBr is expected to exhibit an intralayer ferromagnetic ordering and an interlayer antiferromagnetic ordering, resulting in an A-type antiferromagnetic ground state~\cite{ZiebelNANO-LETT2024}.\newline
A schematic of the samples prepared for this study is shown in Fig.~\ref{fig:1}a. Thin flakes of CrSBr are transferred onto a Au(20 nm)/Ti(5 nm)/SiO$_{x}$/Si substrate via exfoliation in a glove-box under inert nitrogen atmosphere. The sample was degassed at circa 200 $^{\circ}$C (well below the decomposition temperature of CrSBr of 800-850 $^{\circ}$C~\cite{SongJPC-C2025}) for 15 hours in ultra-high vacuum ($\sim 10^{-10}$ mbar) conditions before imaging, in order to remove potential surface contamination deriving from the stacking process and/or the exposure to air during the transfer from the glove-box to the microscope. Figure~\ref{fig:1}b reports a low-energy electron microscopy (LEEM) image of one of the prepared CrSBr flakes on Au/SiO$_{x}$/Si substrate. The inset shows a low-energy electron diffraction (LEED) pattern acquired from the very same CrSBr flake, which allows us to confirm the main in-plane crystallographic directions of the CrSBr flake. The LEEM image shows steps in between different atomic terraces present at the CrSBr surface. The topography of the CrSBr surface is further characterized via atomic force microscopy (AFM) and shown in Fig.~\ref{fig:1}c, offering a more accurate visualization of multiple terraces with different thicknesses.\newline
The magnetic ground state in the CrSBr flake is characterized via low-temperature spin-polarized LEEM (SPLEEM)~\cite{RougemailleEPJAP2010}. The SPLEEM images were acquired after cooling the sample well below the ordering temperature of CrSBr flake via a liquid-He flow cryostat (actual temperature at the sample is circa 30 K). The acquired low-temperature SPLEEM images are shown in Fig.~\ref{fig:1}d-f, which reveal the magnetic contrast along the three main crystallographic directions of the CrSBr flake. The high magnetic contrast in Fig.~\ref{fig:1}e confirms a magnetic easy axis along the in-plane \textit{b}-axis. The weak magnetic contrast in Fig.~\ref{fig:1}d is understood as a small misalignment between the electron beam spin-polarization and the \textit{a}-axis of the CrSBr flake. As LEEM is strongly surface-sensitive, the imaged contrast primarily corresponds to the top CrSBr layer.  
The SPLEEM images reveal the presence of a magnetic multidomain pattern, which does not follow the terraces pattern observed via LEEM (Fig.~\ref{fig:1}b) and AFM (Fig.~\ref{fig:1}c), as expected for an A-type antiferromagnet. These observations clearly point towards a different magnetic ground state hosted by the CrSBr flake, whose nature is discussed in more detail in the following.\newline

\begin{figure}
    \centering
    \includegraphics[width=3.2in]{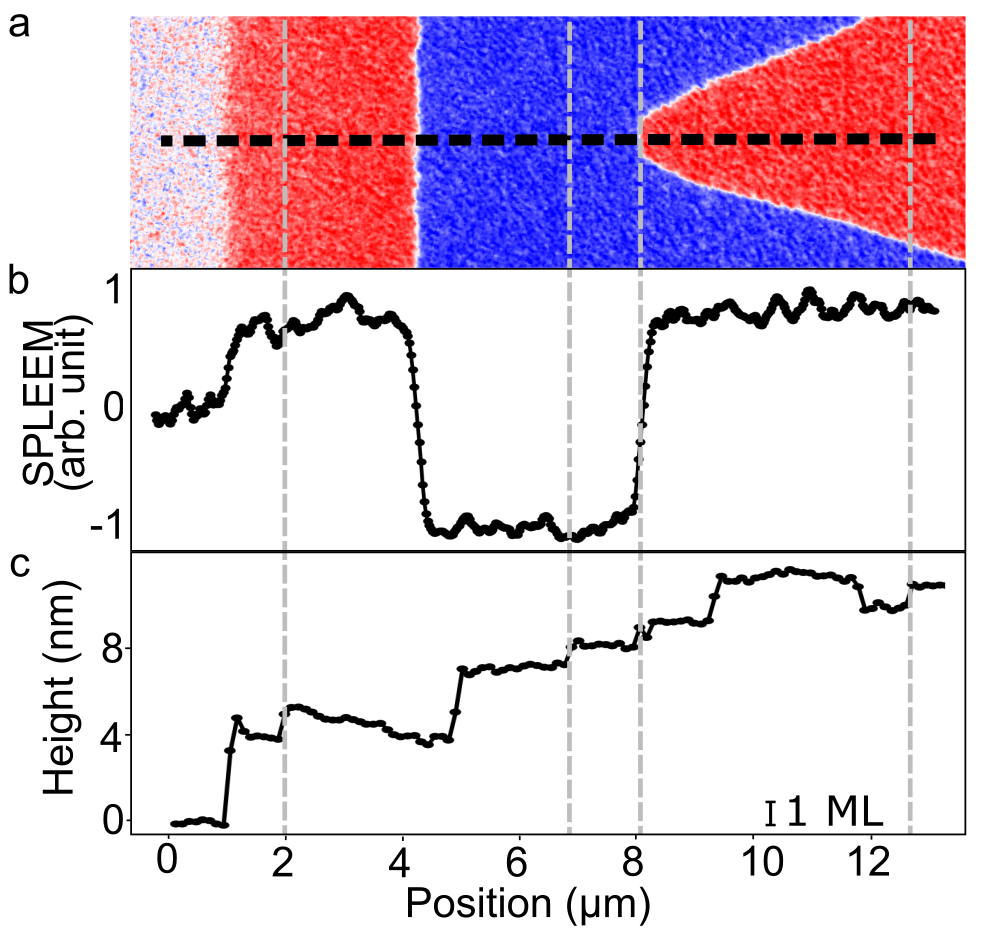}
    \caption{Comparison between observed magnetic contrast and CrSBr thickness. \textbf{a} Zoom-in of the SPLEEM image in Fig.~\ref{fig:1}e, indicating the line along which the cut is taken. \textbf{b} Magnetic contrast along the line displayed in panel \textbf{a}. The contrast is averaged over 10 pixels. \textbf{c} The topographic profiles along the same lines, obtained from the AFM image in Fig.~\ref{fig:1}c. Several of the monolayer steps present in the CrSBr flake, indicated by the vertical grey dotted lines, are found to be located in the middle of ferromagnetic domains.}
    \label{fig:2}
\end{figure}

Figure \ref{fig:2} presents a direct comparison between the magnetic domains present in the CrSBr flake and the topography of the magnetic flake. The SPLEEM image in Figure \ref{fig:2}a shows the observed magnetic domain pattern of the CrSBr flake, with distinct red and blue regions indicating magnetic domains with opposite magnetization. The magnetic domain walls are visible as sharp transition regions between the domains. The dashed black line highlights the line-cut along which the magnetic contrast and the sample's topography are directly compared.
\newline
Figure~\ref{fig:2}b shows the magnetic contrast along the dashed black line indicated in Fig.~\ref{fig:2}a, with a positive/negative magnetic contrast for red/blue magnetic domains, respectively. Finally, Fig.~\ref{fig:2}c shows the CrSBr flake's topography along the same line. This allows for the direct comparison between the observed magnetic contrast and the flake's topography. Importantly, single monolayer (1 ML $\approx$ 0.8 nm)\cite{Telford2022} steps present at the CrSBr surface -indicated in Fig.~\ref{fig:2}c by vertical gray dotted lines- do not correspond to a change in the magnetic contrast shown in Fig.~\ref{fig:2}b. While an inversion of the magnetic contrast is observed at the 1 ML-step in proximity to the apex of the cone-shaped domain, 1 ML-steps elsewhere do not correspond to a magnetization inversion. This observation is directly in conflict with an A-type antiferromagnetic ground state expected for free-standing CrSBr flakes, and demonstrates the emergence of a ferromagnetic interlayer coupling in the studied CrSBr/Au system.

\begin{figure}
    \centering
    \includegraphics[width=6.4in]{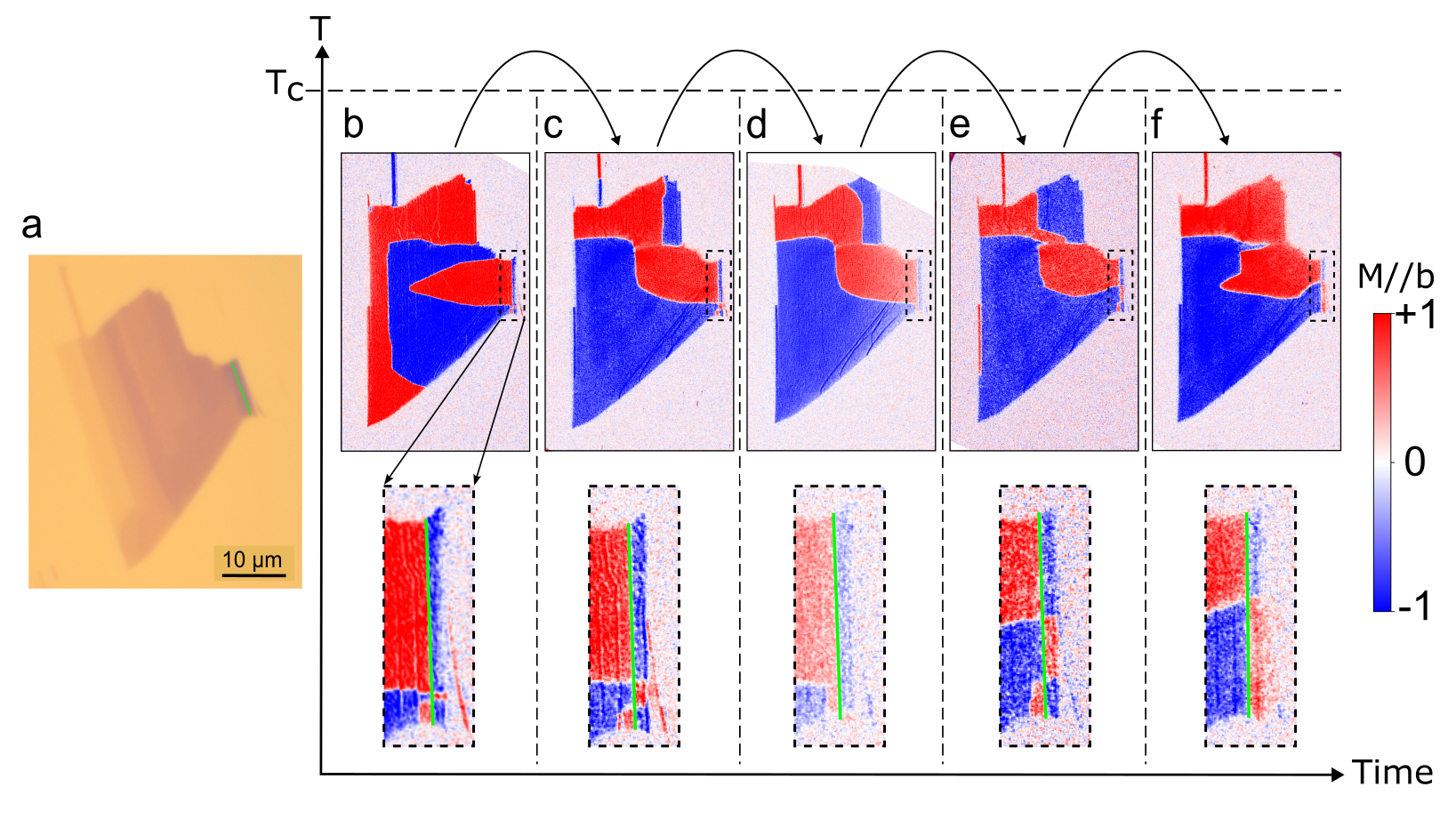}
    \caption{Magnetic domain patterns after warming-cooling cycles. \textbf{a} Optical microscopy image of the CrSBr flake on Au film shown in Fig.~\ref{fig:1}. Lighter (darker) brown regions indicate thinner (thicker) parts of the CrSBr flake. The green line indicates the position of a topographic step on the 2D magnet surface. \textbf{b} SPLEEM image of the magnetic domain pattern in the entire CrSBr flake after the initial cooling down (same as in Fig.~\ref{fig:1}e). \textbf{c}-\textbf{f} SPLEEM images reporting the formed domain patterns after each thermal cycle (warming-up above $T_{C}$ and cooling down below $T_{C}$). A ferromagnetic state in the thin CrSBr flake is formed each time $T<T_{C}$. The bottom raw shows a zoom-in on the thickest part of the flake ($t_{CrSBr}>11$ nm) where an antiferromagnetic interlayer coupling is observed.}
    \label{fig:3}
\end{figure}

To verify that the ferromagnetic ground state reported in Fig.~\ref{fig:2} is not a one-time \textit{freezing} of the particular magnetic domain pattern due to imperfections in the CrSBr flake but rather a general property of the system under investigation, we perform several thermal cycles where we repeatedly warm up the sample above its ordering temperature, $T_{C}\sim132$ K and subsequently cool it down below $T_{C}$, and observe the emerging domain patter after each cycle. The results of this experiment are shown in Fig.~\ref{fig:3}. In Fig.~\ref{fig:3}a there is an optical microscopy image of the same CrSBr flake discussed in Fig.~\ref{fig:1}. Lighter/darker contrast indicates thinner/thicker parts of the 2D magnet. The green line indicates the position of a topographic step. Figure~\ref{fig:3}b shows the initial magnetic domain pattern in the entire CrSBr flake (see top panel). Figure~\ref{fig:3}c-f show the magnetic domain pattern that formed after each warming-cooling cycle. Each SPLEEM image in the top row of Fig.~\ref{fig:3} shows the presence of a magnetic domain pattern which is consistent with the presence of a ferromagnetic order, where ferromagnetic domains are separated by domain walls which do not generally coincide with the 1ML-steps present in the CrSBr flake (as discussed in detail in Fig.~\ref{fig:2}). On the contrary, we observe magnetic domains with same magnetization direction which go across atomic steps of the CrSBr flake. This provides additional evidence for the emergence of a ferromagnetic interlayer exchange in the thin CrSBr flake in direct proximity to the metallic Au film.\newline Importantly, the SPLEEM images at the bottom of Fig.~\ref{fig:3} show a zoom-in on the thickest part of the flake ($t_{CrSBr}>11$ nm) where we observe that the expected interlayer antiferromagnetic coupling of free-standing CrSBr is re-established. The green line indicates the position of the topographic step, where $t_{CrSBr}$ increases from left to right. This observation provides an upper boundary of 11 nm for the thickness across which the induced ferromagnetic ground state remains stable in our system.

\begin{figure}
    \centering
    \includegraphics[width=6.3in]{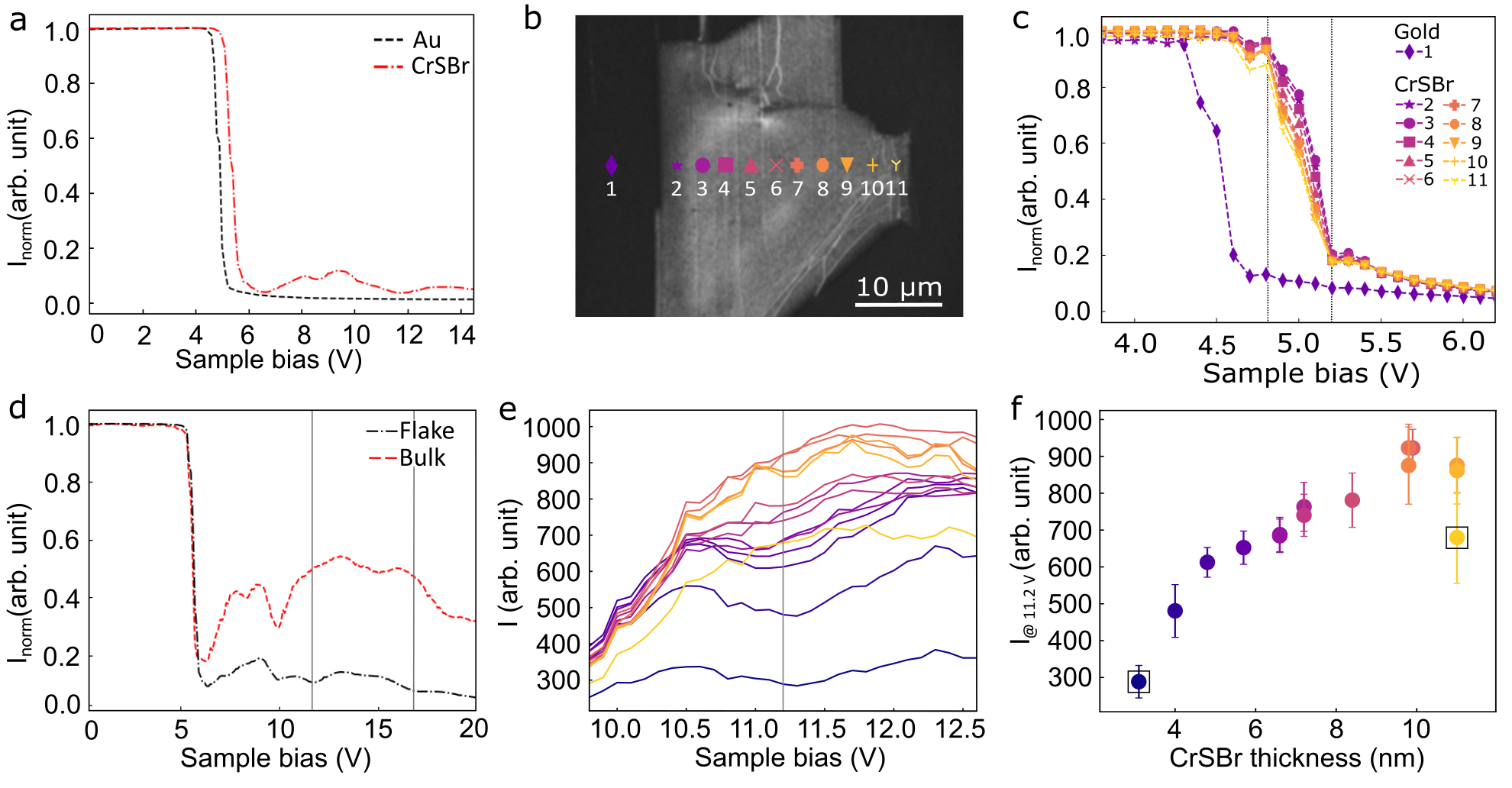}
    \caption{Reflected-electron spectroscopy on the CrSBr($t_{CrSBr}$)/Au system.  \textbf{a} RES spectra showing the reflected electron beam intensity as a function of the sample bias at the polycrystalline gold substrate (grey dashed curve) and at the center of the CrSBr flake (red dotted-dashed curve). \textbf{b} Locations on the sample where the RES spectra shown in panel \textbf{c} are acquired. Two dotted vertical lines in panel \textbf{c} indicate the transition between total electron reflectivity, $V < 4.8$ V and strong electron absorption, $V > 5.2$ V on the CrSBr flake. 
    \textbf{d} RES spectra for bulk CrSBr (thickness $>>$ 10 nm; red dashed curve) and the thin part of the flake of CrSBr (average over $3.5<t_{CrSBr}<8$ nm; black dotted-dashed curve). The gray vertical lines indicate minima in the RES spectrum for the thin CrSBr that are not present in the spectrum of bulk CrSBr. \textbf{f} RES spectra acquired on the CrSBr flake at different thicknesses, going from the left edge (dark blue curve) to the right edge (yellow curve) of the flake shown in panel \textbf{b}. The gray vertical line indicates bias voltage of 11.2 V. \textbf{f} Reflected-electron intensity at a sample bias of 11.2 V, for different CrSBr thicknesses, extracted from RES spectra in panel \textbf{e}. Data points taken closest to the sample edges are highlighted with a black square. Error bars indicate the RMSQ of the RES intensity distributions and of the atomic force microscopy images corrugation. The error bars for the measured thickness are smaller than the symbols.}  
    \label{fig:4}
\end{figure}

To explore the potential origins of the emergent ferromagnetic interlayer exchange coupling in the CrSBr/Au system, we employ spatially-resolved reflected electron spectroscopy (RES), also known as LEEM - \textit{I}(\textit{V}) spectroscopy~\cite{Flege2012}. The aim is to better understand the role played by the interface between the semiconducting CrSBr flake and the metallic Au thin film in the emergent ferromagnetic ground state. 
RES spectra are qualitatively an inverted image of the unoccupied density of states (DOS) - above the vacuum level - of the sample, and in particular they allow us to determine the vacuum level ($E_{vac}$) position. Figure \ref{fig:4}a presents the RES spectra acquired on the CrSBr flake on Au (red dashed-dotted curve), as well as on the Au bare surface (black dashed curve). The distinct reflectivity drop in the RES spectra indicates the energy threshold above which the electrons can penetrate the surface of each material, that is $E_{vac}$; while the features at higher sample biases are linked to the unoccupied DOS at the sample surface. Two main differences can be identified between the RES spectra acquired on CrSBr and Au. Firstly, the RES spectrum acquired on the CrSBr flake has a drop-off at a larger sample bias. Secondly, the RES spectrum acquired on the CrSBr flake has distinctive features for sample biases above the vacuum level, while the RES spectrum from the Au film is featureless. The former observation allows us to state that $E_{vac}$ of the CrSBr flake is approximately 0.4 eV higher than that of the Au thin film, while the latter is a proof of the single crystal nature of the CrSBr flake and of the polycrystalline nature of the Au thin film~\cite{Flege2012}.\newline
Next, we extract RES spectra across the CrSBr flake, going from the thinner (left) to the thicker (right) side of the CrSBr flake, as well as on the Au film, at the locations indicated on the LEEM image in Fig.~\ref{fig:4}b. As the CrSBr thickness increases from approximately 3.5 to 11 nm towards the right side of the flake, any changes in the \textit{I}(\textit{V}) characteristic between these regions can be related to the CrSBr thickness. Fig.~\ref{fig:4}c shows these RES spectra, zoomed in around the drop-off region. Here, the purple diamond curve indicates the spectrum of the Au layer, whereas the other curves indicate spectra taken on CrSBr of increasing thickness. The spectra taken on CrSBr all show similar behavior. The reflected intensity starts dropping quickly around a bias voltage of 4.8 V (see vertical dotted line on the left), after which it flattens out at a bias voltage of 5.2 V (see vertical dotted line on the right). Accordingly, within the energy resolution of the instrument (e-beam energy spread $\approx$ 200 meV~\cite{DaleARXIV2024}), we do not observe a clear shift of the vacuum level of CrSBr as a function of its thickness, for $3.5 \leq t_{CrSBr} \leq 11$ nm.\newline The difference between the extracted $E_{vac}$ on CrSBr and Au indicates the presence of a built-in potential, $V_{bi}$, at the interface. $V_{bi}$ results in the redistribution of charges between Au and CrSBr, forming an electron accumulation region in the semiconductor. By solving Poisson's equation within a full depletion approximation, the dopant density can be expressed in terms of $V_{bi}$ and the width of the accumulation layer $W$ as $N_{D} = \frac{2\epsilon_0\epsilon_rV_{bi}}{qW^2}$. As no thickness-dependent band-bending is observed, an upper limit of 3.5 nm is taken for $W$. Using an approximate value of 10 for $\epsilon_r$ \cite{Wang2023}, this yields a minimum electron doping density of $N_D \approx 10^{19}$ cm$^{-3}$, corresponding to a sheet density of about $10^{13}$ cm$^{-2}$. \newline
The unoccupied DOS of the CrSBr flake is further investigated in Fig.~\ref{fig:4}d, where the RES spectrum is plotted for two different thicknesses of CrSBr. The black dashed-dotted curve shows the average spectrum taken on the thin part of the CrSBr flake in Fig.~\ref{fig:4}b ($3.5<t_{CrSBr}<8$ nm), whereas the red dashed line corresponds to a bulk CrSBr sample ($t_{CrSBr}>> 10$ nm). Above $E_{vac}$, both curves show several features corresponding to the unoccupied DOS of the material. Minima (high DOS) and maxima (low DOS) in the two curves occur at different sample biases, indicating that the electronic structure above the vacuum level is different for the bulk and the thin flake. In particular, two minima at bias voltages equal to 11.2 V and 16.4 V (see vertical gray lines in Fig.~\ref{fig:4}d) present in the RES spectrum acquired on thin CrSBr on Au are not present in the spectrum acquired on the bulk CrSBr crystal. The spectral feature at sample bias = 11.2 V is further investigated in terms of the CrSBr thickness. Fig.~\ref{fig:4}e shows 14 RES spectra acquired across the CrSBr flake shown in Fig.~\ref{fig:4}b from the left edge to the with edge, in the 9.8 V - 12.6 V bias voltage range. A clear thickness dependency in the \textit{I}(\textit{V}) spectra can be observed, where the minimum at sample bias = 11.2 V visible for the thinnest part (blue curves) of the CrSBr flake is not observed for the thicker part (orange-yellow curves) of the flake, transitioning towards a more bulk-like character (see dashed red curve in Fig.~\ref{fig:4}d). Finally, Fig.~\ref{fig:4}f shows the reflected-electron intensity at sample bias = 11.2 V as a function of $t_{CrSBr}$, further highlighting the evolution of such spectral feature in the thickness range of the investigated CrSBr flake. It is worth noting that the two data points in  Fig.~\ref{fig:4}f  coming from locations on the CrSBr surface in proximity to the edges  show a strongly reduced RES intensity (see data points enclosed in black squares). This is understood as a measurement artifact due to the semiconducting nature of the 2D magnet, which results in the distortion of the electric field lines at the sample's edges. Nonetheless, the rest of the data points in Fig.~\ref{fig:4}f show an evident thickness dependency. This is experimental evidence of thickness-dependent band renormalization in the thin CrSBr flake~\cite{Ghimirey-npj2DMAT_APPL2026}, which can potentially originate from interface effects such as band hybridization~\cite{Bianchi2023-kx} and/or screening. \newline

\begin{figure}
    \centering
    \includegraphics[width=3.2in]{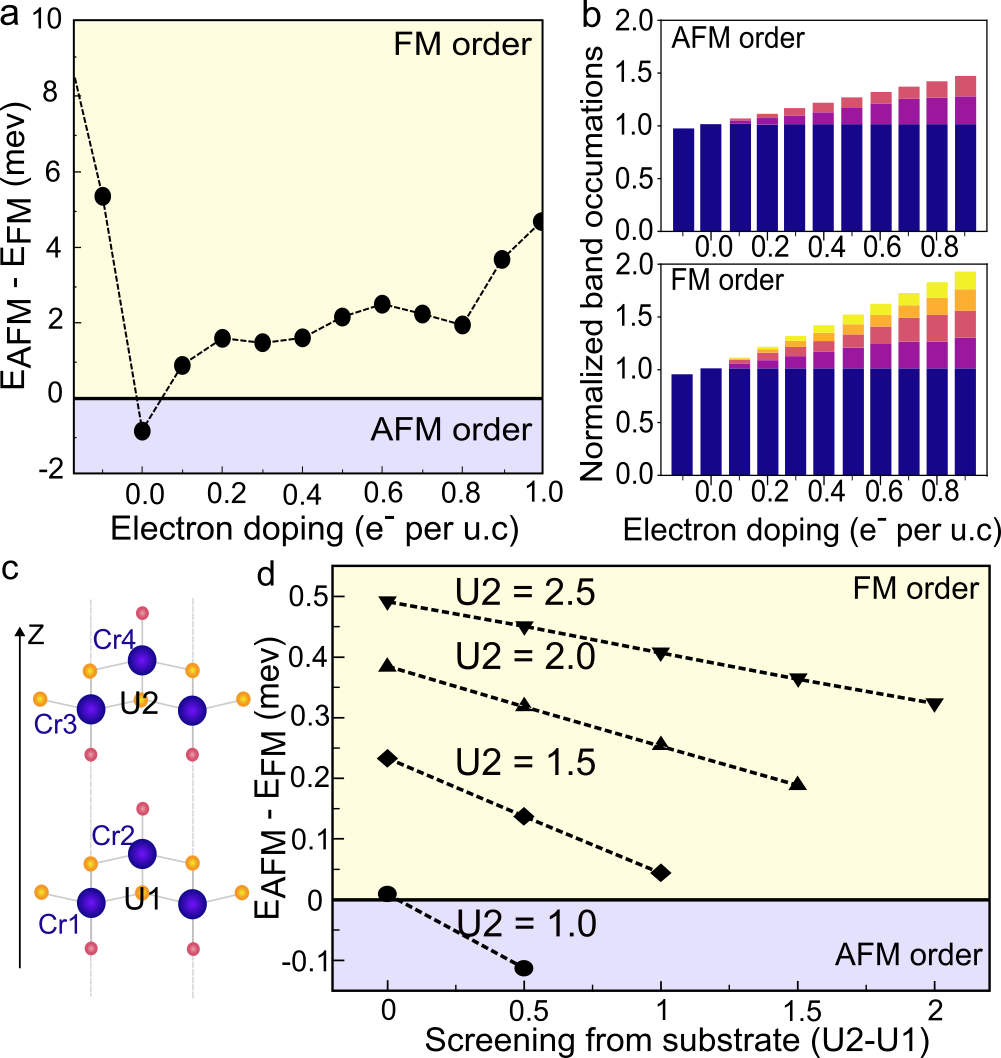}
    \caption{Computed dependence of interlayer exchange coupling in CrSBr with respect to electron doping and substrate screening. \textbf{a} The computed total energy difference between the antiferromagnetic (AFM) and the ferromagnetic (FM) stacking in bulk CrSBr. Both positive and negative electron doping are found to promote the emergence of an interlayer FM alignment over the AFM one expected for a free-standing CrSBr flake. \textbf{b} The band occupation normalized to the valence band occupation at zero doping as a function of the electron doping for both a fixed AFM (top) and FM (bottom) ordering, respectively. The blue bars refer to the valence band occupation, whereas the purple, red, orange and yellow bars correspond to different conduction bands. \textbf{c} Schematic of a CrSBr bilayer with different local Coulomb interaction, \textit{U}, for top and bottom layer used to explore the effect of electrostatic screening from the Au metallic layer on the interlayer exchange interaction. \textbf{d} The computed total energy difference between the AFM and the FM stacking in bilayer CrSBr as a function of \textit{U}$_{2}$-\textit{U}$_{1}$. A gradient in \textit{U} is always found to promote an AFM ordering against a FM one, while a large \textit{U} promotes a FM ordering.}
    \label{fig:5}
\end{figure}

To investigate the potential origin of the observed ferromagnetic interlayer exchange interaction in the thin CrSBr flakes in proximity to a metallic film, we perform first principles density functional theory (DFT) calculations. Strong electron doping has been previously observed in a few layer CrSBr on a gold substrate \cite{Bianchi2023-kx}. As a similar sample structure is employed here, electron doping is expected to be present. To investigate the influence of such doping on the magnetic ground state of the system, we first calculate for bulk structures the difference of the total energies of the ferro- and antiferromagnetic interlayer ground states, denoted by $E_{FM}$ and $E_{AFM}$, respectively, as a function of electron and hole doping. As shown in Fig.~\ref{fig:5}a, at zero doping the interlayer antiferromagnetic state is energetically slightly more favorable, as expected for free-standing CrSBr~\cite{ZiebelNANO-LETT2024}. However, both electron and hole doping are found to promote the emergence of a ferromagnetic interlayer coupling over the antiferromagnetic alignment.\newline
Even though the ferromagnetic arrangement is observed to become more and more energetically favorable for increasing electron doping, the trend is not entirely continuous. This results from the subsequent occupation of the conduction bands, which is different in the AFM and FM order. As shown in Fig.~\ref{fig:5}b, in the case of fixed anti-ferromagnetic interlayer coupling, the doping electrons occupy only two conduction bands (purple and red bars). While, in the fixed ferromagnetic state, there are up to four conduction bands that get partially occupied (purple, red, orange and yellow bands). On the hole doping side, we find the FM state energetically much more favored, with an (initially) more homogeneous trend in $E_{AFM}-E_{FM}$ which is accompanied by a similar de-population of the same highest valence bands in both interlayer coupling calculations.\newline The outcome of the first-principles DFT calculations speaks for a possible role played by the Fermi surface in the definition of the magnetic properties of CrSBr flakes with electron/hole doping. One possible scenario which could explain the asymmetric response to the type of doping, is that of a carrier-mediated Ruderman–Kittel–Kasuya–Yosida (RKKY) interaction, favoring the FM state. Since the Fermi surfaces at hole and electron doping are rather different \cite{Klein2023}, the RKKY interaction is expected to be rather different in the two cases, as it depends on both the effective mass and the Fermi wave vector. A second possible scenario is that of a super-exchange mechanism between the neighboring layers being affected by electron/hole doping, as it results from virtual hopping from occupied to unoccupied sites. Thus, if the occupation of certain states changes upon doping, some virtual hoppings might be suppressed, affecting the super-exchange mechanism.\newline
A second aspect which is considered as a potential drive for the observed ferromagnetic ground state is electrostatic screening. Indeed, having a thin layer of CrSBr in direct proximity to a metallic Au layer naturally raises the question of whether the Coulomb interactions between the charge carriers in the thin magnetic semiconductor are strongly affected by the presence of a metallic layer underneath. For the discussion of interlayer magnetic coupling, we specifically investigate here how a certain gradient in the Coulomb interactions strength (in the $z$-direction, perpendicular to the CrSBr-Au interface) might affect the interlayer magnetic coupling. This kind of gradient is to be expected, as the screening from the Au layer diminishes with an increasing distance from the CrSBr-Au interface. Accordingly, we explore the effect of such a gradient in the Coulomb potential, $\Delta$\textit{U} = $U_{2}-U_{1}$, between the layers of bilayer CrSBr (see schematic in Fig.~\ref{fig:5}c), with $U_{1} (\leq U_{2})$ beeing the effective Coulomb repulsion in the CrSBr layer supposedly closer to the Au substrate. The results of the calculations reported in Fig.~\ref{fig:4}d show that increasing $\Delta$\textit{U} has a tendency towards stabilizing an AFM coupling, as $\Delta$\textit{E}($\Delta$\textit{U}) = $E_{AFM}-E_{FM}$ decreases with $\Delta$\textit{U}; while large values of \textit{U} favor a FM interlayer coupling.\newline The two analyses above show that $\Delta$\textit{E} is much more sensitive to the electron doping than to $\Delta$\textit{U}. Based on this comparison between electron doping and screening (gradient) trends, we can conclude that electron doping is expected to be one of the possible drives behind the emergence of the observed ferromagnetic interlayer coupling.

\section{Discussion}
Our experimental analysis shows that individual layers in thin CrSBr flakes on Au, for a CrSBr thickness smaller than 11 nm, couple ferromagnetically with each other. This is accompanied by the observation of band renormalization. On the theoretical side, it was found before that the electronic band structure of CrSBr changes with the magnetic interlayer order~\cite{BianchiPRB2023,WatsonNPJ-2DmaterAppl2024,YuADV-FUNC-MATER2024,ZhaoNAT-COMMUN2025}. Here we completement this by ab-initio calculations showing that the stability of the interlayer antiferromagnetic coupling in pristine CrSBr is weak and that small perturbations, e.g. doping due to electron/hole transfer, can easily tip the system towards a ferromagnetic ground-state. Our detailed analysis furthermore showed that a gradient in the local Coulomb interactions between neighbouring layers, as it might be induced from screening by the metallic Au film, has a (mild) tendency towards stabilizing an AFM coupling. As no thickness dependency is observed for the vacuum level, $E_{vac}$, of the thin CrSBr flake, within the $\approx$ 200 meV experimental resolution, the expected electron doping from the Au substrate must be localized near the CrSBr-Au interface, with a lower bound for the doping concentration estimated to be of the order of $N\approx 10^{19}$ cm$^{-3}$, or approximately 0.1 e$^{-}$/u.c. This doping concentration is shown to promote ferromagnetic coupling in the CrSBr in our calculations. It is worth stressing that the AFM order in CrSBr is expected to be weak, in agreement with previous estimates of small interlayer magnetic couplings for pristine CrSBr~\cite{denTeulingPRB2025} and so prone to be influenced by external stimuli, as demonstrated by the observed strong responses to rather weak magnetic fields~\cite{YuADV-FUNC-MATER2024}, gating~\cite{ZhaoNAT-COMMUN2025,HongNAT-COMMUN2025}, or stacking~\cite{MondalARXIV2025}.\newline
A potential additional force behind the stabilization of the observed ferromagnetic phase is strain generated into the CrSBr flake by local bending during the exfoliation process. It was reported that the tuning of the in-plane lattice constant in CrSBr could realize a ferromagnetic phase~\cite{CenkerNAT-NANO2022}. However, we do not have any experimental evidence that connects the observed ferromagnetic phase with strain in our study. The strain induced in the exfoliated CrSBr is rather non-homogeneous, most likely varying locally from tensile to compressive across the flake and with no specific periodicity, in contrast with the uniform uni-directional strain considered in the study by Cenker at al.~\cite{CenkerNAT-NANO2022}. All this seems to role out strain as the main cause of the observed ferromagnetic phase in our system.\newline In conclusion, given the fact that charge transfer into the CrSBr flake is the most likely origin of the observed ferromagnetic phase, our findings suggest that the behaviour of CrSBr-based devices with direct contact to metallic leads needs to be very carefully interpreted in terms of their interlayer magnetic coupling properties in vicinity of the contacts. Finally, our study highlights the utility of directly imaging the magnetic ground state of thin van der Waals materials via SPLEEM and sheds new light on the possibility to engineer the magnetic properties of 2D van der Waals magnets via proximity/interface effects with metallic layers.

\section*{Author contributions}
$^{\times}$ authors contributed equally. \newline
R.L.C. conceived the study. R.B. prepared the samples under M.H.D.G.'s supervision. R.L.C., N.D., and C.W.J. performed the experiments at the SPLEEM Lab at LBNL under A.S.'s supervision. D.A.P., D.S., J.W. developed the data analysis code and analyzed the data with the support of R.B. and R.L.C.. S.G., A.R., and M.R. performed the first principles DFT calculations. R.B., D.A.P., and  R.L.C. wrote the manuscript. R.L.C. coordinated the writing of the manuscript. All authors commented on the manuscript.

\section*{Acknowledgments}
R.L.C., R.B. and M.H.D.G. acknowledge financial support from the Zernike Institute for Advanced Materials of the University of Groningen. R.B. and M.H.D.G. acknowledge financial support from the Dutch Research Council (NWO, OCENW.XL21.XL21.058) and the European Research Council (ERC, 2D-OPTOSPIN, 101076932). D.A.P. acknowledges financial support from the Top Master Programme in Nanoscience at the Zernike Institute for Advanced Materials of the University of Groningen. S.G. and M.R. acknowledge support from the
Vidi ENW research program of the Dutch research council (NWO) [Grant DOI: 10.61686/YDRHT18202] with File No. VI.Vidi.233.077. Sample fabrication was performed using NanolabNL facilities. The work at the Molecular Foundry of the Lawrence Berkeley National Laboratory was supported by the Office of Science, Office of Basic Energy Sciences, of the U.S. Department of Energy under Contract No. DE-AC02-05CH11231, under user proposal numbers 08459 and 10102.

\newpage

\section*{Supplementary Information}
\subsection*{Sample fabrication}
SiO$_x$/Si substrates are cleaned in acetone. 5 nm of titanium and 20 nm of gold is deposited onto the substrate via electron beam evaporation. Bulk CrSBr crystals were purchased for HQ graphene and exfoliated in a nitrogen-filled filled glove-box using mechanical exfoliation onto a commercial PDMS film (gel pak). Suitable flakes were selected based on optical contrast and transferred onto the gold-coated SiO$_x$/Si substrate.
\subsection*{Low-temperature magnetic imaging via SPLEEM}
The magnetic imaging of the CrSBr flakes on Au/SiO$_x$/Si substrates, as well as the reflected-electron spectroscopy measurements, were conducted with the low-temperature SPLEEM system (Elmitec Model LEEM III with spin-polarized electron beam source, spin-manipulator and custom-designed liquid helium cryostage) available at the Molecular Foundry of the Lawrence Berkeley National Laboratory. The system allows for the acquisition of images via low-energy electrons with magnetic contrast by employing a spin-polarized electron beam as the illumination source. The spin orientation of the electron beam can be selected along any desired direction in three-dimensional space. The SPLEEM images were acquired with samples cooled down via a liquid-He flow cryostat. The magnetic contrast is a direct result of the different reflection probability for spin-up and spin-down electrons at the magnetic surface. The origin of such spin-dependent electron reflection probability is the spin-polarized density of states of the magnetic surface. The magnetic asymmetry in a SPLEEM image is calculated as $A =[ (S_{u} - S_{d})/(S_{u} + S_{d})]$, where $S_{u}$ and $S_{d}$ are the backscattering intensity of the spin-up and spin-down polarized electron-beam, respectively. Images are drift-corrected before performing averaging in order to maximize the quality of the presented SPLEEM images. The drift correction is performed via cross-correlating all the LEEM images contained in the acquired image stack.
\subsection*{Reflected-electron spectroscopy}
During the acquisition of reflected-electron spectra, the landing energy of the electrons is controlled via the application of a bias voltage (starting voltage) between sample and cathode. Reflected-electron spectroscopy is conducted by varying the landing energy of the electrons from a low value to a high value. In the low energy regime, the electrons energy is lower than the vacuum level of the sample and are fully reflected (mirror mode). At a particular applied bias voltage, the electrons have enough energy to penetrate the sample surface and probe the unoccupied energy levels above the vacuum level at the sample's surface (LEEM mode). A custom-developed Python code was employed to extract spatially resolved reflected-electron spectra from the acquired LEEM(V) image stacks.
\subsection*{DFT calculations}
All DFT calculations are performed within the projected augmented wave formalism \cite{paw} as implemented in the Vienna Ab-Initio Simulation Package~\cite{Kresse1993, Kresse1996} (VASP). The exchange-correlation effects were considered within the GGA-PBE functional \cite{pbe}. To account for the on-site Coulomb repulsion within the 3$d$ shell of Cr atoms, we used the standard DFT+$U$ scheme \cite{ldapu} with $U=4\,$eV and $J=1\,$eV. To ensure sufficient numerical accuracy of our results, we apply a (16$\times$12$\times$4) $k$-point grid for Brillouin zone integrations, and an energy cut-off of 800 eV for the plane waves. The lattice constants and atomic positions are taken from the Materials Project for orthorhombic CrSBr (mp-22998), and then fully relaxed using the parameters of our calculations at charge neutrality, yielding $a\approx3.45\,$\AA, $b\approx4.66\,$\AA\ and $c\approx15.56\,$\AA. To capture the van der Waals (vdW) dispersion forces between the CrSBr layers during the relaxation, we apply a vdW correction within the DFT-D3 method \cite{Grimme2011}. We note that to stabilize the interlayer AFM ground state at charge neutrality both, applying $J=1\,$eV and the full relaxations were important. Thus, the lattice structure and the applied Coulomb parameters differ slightly from previously used ones~\cite{Rudenko2023}.

\newpage

\bibliography{sn-bibliography}

\end{document}